\documentclass[12pt]{article}
\usepackage{graphicx} 
\usepackage{xcolor}
\usepackage{amsmath}
\usepackage{algorithm}
\usepackage{amsmath}
\usepackage{amssymb}
\usepackage{array}
\usepackage{indentfirst}
\usepackage{algpseudocode}
\usepackage{booktabs}
\usepackage{float}
\usepackage{tabularx}
\usepackage{float}
\usepackage{caption}
\providecommand{\keywords}[1]
{
  \small	
  \textbf{Keywords:} #1
}

\title{Dynamic Investment Strategies Through Market Classification and Volatility: A Machine Learning Approach\thanks{Research supported by ADIALab}}

\author{
Jinhui Li\thanks{Department of Mathematics, University of Toronto, Toronto, ON, Canada, M5T 3J1. E-mail: jinhuidavis.li@mail.utoronto.ca},
Wenjia Xie\thanks{Department of Mathematical Sciences, Tsinghua University, Beijing, China, 100084. E-mail: jiewj20@mails.tsinghua.edu.cn},
Luis Seco\thanks{Department of Mathematics, University of Toronto, Toronto, ON, Canada, M5T 3J1. E-mail: luis.seco@utoronto.ca}
}

\date{March 2025}

\begin{document}

\maketitle
\begin{abstract}
This study introduces a dynamic investment framework to enhance portfolio management in volatile markets, offering clear advantages over traditional static strategies. It evaluates four conventional approaches—equal weighted, minimum variance, maximum diversification, and equal risk contribution—under dynamic conditions. Using K-means clustering, the market is segmented into ten volatility-based states, with transitions forecasted by a Bayesian Markov switching model employing Dirichlet priors and Gibbs sampling. This enables real-time asset allocation adjustments. Tested across two asset sets, the dynamic portfolio consistently achieves significantly higher risk-adjusted returns and substantially greater total returns, outperforming most static methods. By integrating classical optimization with machine learning and Bayesian techniques, this research provides a robust strategy for optimizing investment outcomes in unpredictable market environments.

\end{abstract}

\keywords{
Machine Learning, Dynamic Portfolio Allocation, Market Classification, Bayesian Markov Methods, Dirichlet Priors and Gibbs Sampling
}

\section{Introduction}

The seminal work of Markowitz (1952)~\cite{markowitz1952} established the foundation for modern portfolio theory, emphasizing the importance of diversification and optimizing asset allocation to enhance returns and manage risks. However, financial markets are inherently volatile and constantly changing, making static portfolio strategies less effective over time. The problem we are tackling is how to optimally adjust portfolio allocations in response to these dynamic market conditions. This problem is important because failure to adapt to market volatility can result in suboptimal returns and increased risk exposure. Recent developments in machine learning and statistical modeling have opened new avenues for advancing these strategies, particularly through more sophisticated analysis of market states and volatility. We aim to solve the optimal portfolio selection problem by addressing it within different volatile states, thus enhancing the robustness and performance of investment strategies. 

Traditional asset allocation strategies, such as equally-weighted investment, minimum variance, equal risk contribution, and maximum diversification, have shown varying degrees of effectiveness in managing portfolio risks and returns~\cite{demiguel2009, maillard2010}. For example, DeMiguel et al. (2009)~\cite{demiguel2009} showed that simple allocation strategies often outperform more complex optimization-based strategies, while Maillard et al. (2010)~\cite{maillard2010} proposed the concept of risk parity to balance the risk contribution of each asset in a portfolio. However, these strategies often assume static market conditions or rely on retrospective data, limiting their adaptability to sudden changes in market volatility. More recent approaches, such as those of Kritzman et al. (2010)~\cite{kritzman2010}, have begun to explore dynamic strategies that adapt to changing market conditions. This research introduces a dynamic approach to investment strategy, utilizing a machine learning-based clustering method to categorize market states into distinct segments based on historical market returns and volatilities.

Traditional methods often fail to adapt quickly to sudden market changes, as they are typically based on historical data without consideration for evolving market conditions. For instance, strategies like minimum variance or equally-weighted investment assume that past data can reliably predict future risks and returns, which is not always the case during market upheavals. This limitation is crucial, as it can lead to suboptimal asset allocation and increased risk during periods of market turbulence. As highlighted by Kritzman et al. (2010)~\cite{kritzman2010}, dynamic strategies that account for changing market states can provide better risk management and return optimization. Similarly, Escobar et al. (2013)~\cite{escobar2013market} emphasize the limitations of static allocation strategies such as the 1/N approach, which equally weights assets regardless of their risk-return profiles, potentially leading to inefficiencies during market crises. Dynamic strategies, therefore, offer a more responsive approach to asset allocation, adapting to market conditions, and reducing risk more effectively than traditional static methods.

In this study, we address the limitations of traditional asset allocation strategies by introducing a dynamic approach to investment. The market is initially divided into ten states using the K-means clustering algorithm, which allows for a detailed exploration of the interaction between market states and investment performance. This segmentation helps us to understand how different market conditions affect the efficacy of various portfolio strategies. Following the classification, we test four traditional portfolio methods—equally weighted investment, minimum variance, equal risk contribution, and maximum diversification—along with a dynamic portfolio method across these ten states. Our goal is to identify which portfolio method yields the highest return, lowest volatility, and best risk-adjusted return (Sharpe ratio) in each state. The best-performing portfolio method for each state is then used to construct a dynamic portfolio that adapts to changing market conditions.

To ensure the robustness of this approach, we thoroughly address data quality and preprocessing. Comprehensive performance metrics are incorporated into the analysis, including annual return, annualized volatility, and the Sharpe ratio. The accuracy of the clustering and the correctness of the state assignments are validated through rigorous statistical techniques. This validation process ensures that the results are reliable and can be generalized to various market conditions, thereby increasing the applicability and trustworthiness of the findings.
Subsequently, a Bayesian Markov switching model is employed to navigate and capitalize on the dynamic nature of these states. The transition probabilities between states are calculated using Dirichlet prior and Bayesian Markov Chain Monte Carlo (MCMC) methods, specifically Gibbs sampling. This facilitates a more robust analysis of state transitions and probabilistic forecasting of market conditions. These probabilities are used to dynamically adjust asset allocation strategies, building a final portfolio method that adapts to real-time market conditions.

The ultimate goal of this research is to identify the optimal investment strategy for each volatility-defined market state and dynamically adjust the portfolio based on the Markov transition probabilities, thus optimizing the decision-making process in real-time market conditions. This analysis reveals that the dynamic portfolio strategy, based on return weights, significantly improves risk-adjusted returns and reduces volatility compared to static strategies. For the first asset, the dynamic portfolio consistently outperforms all other methods except for the ERC strategy in terms of return and Sharpe ratio. For the second asset, the dynamic portfolio outperforms all methods, including ERC. This paper extends the existing body of knowledge by integrating classical financial theories with cutting-edge machine learning techniques to create a more responsive and effective portfolio management framework. Through empirical analysis and model testing, this study demonstrates that a more granular understanding of volatility-driven market states can significantly enhance the robustness and performance of investment strategies~\cite{kritzman2010}.

The remainder of the paper is organized as follows. In Sect.\ref{sec:method}, we discuss and present our proposed methodology, including the market state classification using K-means clustering and the construction of the Bayesian Markov transition matrix. Sect.\ref{sec:empirical} details the empirical implementation and results, showcasing the performance of the dynamic portfolio strategy compared to static methods. In Sect.\ref{sec:discussion}, we provide a discussion on the findings and their implications. Finally, Sect.\ref{sec:conclusion} concludes the article, highlighting key contributions and suggesting avenues for future research.

\section{Methodology}\label{sec:method}


\subsection{K-Means Clustering for Market Segmentation}

To capture distinct market conditions, we segment historical volatility into ten states using K-means clustering. This algorithm groups observations by minimizing the within-cluster sum of squares, effectively identifying regimes of low, medium, and high volatility. We apply it to 22-day volatility data, enabling a granular analysis of how portfolio strategies perform across different market states. The process starts by randomly initializing ten centroids, iteratively assigning data points to the nearest centroid, and updating centroids until convergence. This classification forms the foundation for our dynamic allocation strategy.

The K-means algorithm partitions \( n \) observations into \( k \) clusters, where each observation belongs to the cluster with the nearest mean. The objective is to minimize the within-cluster sum of squares (WCSS), defined as:

\begin{equation}
\text{WCSS} = \sum_{i=1}^{k} \sum_{x \in C_i} \| x - \mu_i \|^2,
\end{equation}

where \( C_i \) is the set of observations in cluster \( i \) and \( \mu_i \) is the mean of the observations in cluster \( i \).

The K-means clustering procedure involves the following steps:

1. \textbf{Initialization}: Randomly select \( k \) initial cluster centroids.

2. \textbf{Assignment}: Assign each observation to the nearest centroid based on the Euclidean distance.

3. \textbf{Update}: Calculate the new centroids as the mean of the observations assigned to each cluster.

4. \textbf{Repeat}: Repeat the assignment and update steps until the centroids converge (i.e. their positions no longer change).

Mathematically, the assignment step can be represented as:

\begin{equation}
C_i = \{ x_p : \| x_p - \mu_i \|^2 \leq \| x_p - \mu_j \|^2 \text{ for all } j, 1 \leq j \leq k \},
\end{equation}

The update step is then:

\begin{equation}
\mu_i = \frac{1}{|C_i|} \sum_{x_j \in C_i} x_j.
\end{equation}

\subsection{Bayesian Markov Switching Model Using Bayesian Transition Matrix}

To model the transitions between the identified states, we employed a Bayesian approach to estimate the transition probabilities. This method incorporates prior knowledge through the use of a Dirichlet prior and leverages Markov Chain Monte Carlo (MCMC) methods, specifically Gibbs sampling, to derive the transition probabilities. This approach provides a robust probabilistic framework that can adapt to the uncertainty inherent in the data.

In constructing the Bayesian transition matrix \( P \) for a Markov chain with states \( 1 \) through \( 10 \), We begin by initializing the Dirichlet prior, chosen for its conjugate properties with the Multinomial distribution and its suitability for modeling probability vectors that must sum to 1. The concentration parameters for the Dirichlet distribution, representing prior counts for transitions from each Markov state, are denoted as:

\begin{equation}
\alpha_{i} = (\alpha_{i1}, \alpha_{i2}, \dots, \alpha_{i10})
\end{equation}

Here, \( \alpha_{ij} \) represents the prior belief (or prior counts) about the transition from Markov state \( i \) to state \( j \).

Next, we count the number of posterior transitions from each state \( i \) to every other state \( j \):

\begin{equation}
N_{ij} = \text{Number of posterior transitions from state } i \text{ to state } j
\end{equation}

The posterior distribution for the transition probabilities is given by the Dirichlet distribution, which is the conjugate prior of the multinomial distribution. Thus by the Bayesian theorem, For the vector of transition probabilities \( P_i \) (from state \( i \) to all other states), the posterior is:

\begin{equation}
P_i = (P_{i1}, P_{i2}, \dots, P_{i10}) \sim \text{Dirichlet}(\alpha_{i1} + N_{i1}, \alpha_{i2} + N_{i2}, \dots, \alpha_{i10} + N_{i10}),
\end{equation}

where we have \begin{equation}
E[P_{ij}] = \frac{\alpha_{ij} + N_{ij}}{\sum_{k=1}^{10} (\alpha_{ik} + N_{ik})}.
\end{equation}

This process is repeated for a sufficiently large number of iterations to ensure convergence, which can be assessed using the Gelman-Rubin Diagnostic\cite{gelman1992inference} with a threshold of $1.1$.

In practice, updates to the transition matrix \( N_{ij}^{(k)} \) will occur once per period, such as every three or six months. Therefore, \( N_{ij}^{(k)} \) forms a sequence that updates the posterior distribution repeatedly , as shown in (3.6). This sequential updating process is the primary reason we apply Gibbs sampling rather than assuming the transition matrix follows a fixed Dirichlet distribution.

When updating the posterior row-by-row with the Dirichlet distribution \( \text{Dirichlet}(\boldsymbol{\alpha}_i + \sum_{j=1}^{k} \boldsymbol{N}_i^j) \), where \( \boldsymbol{\alpha}_i \) is the prior vector for state \(i\), and \( \boldsymbol{N}_i^j \) is the count vector at time \(j\) for state \(i\), the model reflects the cumulative data. Gibbs sampling iteratively refines this posterior by retaining past information, integrating it over time, and converging to the true probability model, ensuring a comprehensive understanding of both past and current data.

This Bayesian approach, using the Dirichlet prior and Gibbs sampling, provides a flexible and robust method for estimating transition probabilities, accommodating the inherent uncertainty and variability in the data. By assuming that each \( P_i \) (the transition probabilities for each state) is independent, the conditional distribution in Gibbs sampling becomes simple, as it only requires updating one \( P_i \) at a time while conditioning on the others. 

\subsection{Mixing Time}

Mixing time is a crucial concept in the analysis of Markov chains, particularly when using MCMC methods like Gibbs sampling to estimate transition matrices.

The mixing time of a Markov chain is the time it takes for the chain to converge to its stationary distribution. Mathematically, it is defined as the smallest \(t\) such that the total variation distance between the distribution at time \(t\) and the stationary distribution \(\pi\) is less than a threshold \(\epsilon\):

\begin{equation}
t_{\text{mix}}(\epsilon) = \min \left\{ t \mid \max_{x} \| P^t(x, \cdot) - \pi(\cdot) \|_{\text{TV}} \leq \epsilon \right\},
\end{equation}

where \(\| \cdot \|_{\text{TV}}\) denotes the total variation distance and \(P^t(x, \cdot)\) is the distribution of the chain at time \(t\) starting from state \(x\).

To calculate the mixing time of a Markov chain, we often use bounds based on the eigenvalues of the transition matrix \(P\). For an irreducible, reversible and aperiodic Markov chain, the mixing time can be bounded using the second-largest eigenvalue modulus (SLEM), as follows \cite{levin2009markov}:

\begin{equation}
(t_{\text{rel}} - 1) \log \left( \frac{1}{2\epsilon} \right) \leq t_{\text{mix}}(\epsilon) \leq t_{\text{rel}} \left( \frac{1}{2} \log \left(\frac{1}{\pi_{\min}}\right) + \log \left(\frac{1}{2\epsilon}\right) \right)
\end{equation}

where \( t_{\text{rel}} = \frac{1}{1 - \lambda_2} \) is the relaxation time, \(\lambda_2\) is the second-largest eigenvalue modulus of the transition matrix \(P\), and \(\pi_{\min} = \min_{x \in X} \pi(x)\) is the minimum probability in the stationary distribution. Note that for a reversible Markov transition matrix, the largest eigenvalue is always $1$, and all eigenvalues are real numbers bounded between $-1$ and $1$.

 These bounds indicate that the convergence rate of the Markov chain depends on how close \(\lambda_2\) is to 1. A smaller spectral gap (\(1 - \lambda_2\)) implies slower convergence, as the chain takes longer to mix. The parameter \(\pi_{\min}\) also affects the upper bound, reflecting the influence of the least probable state in the stationary distribution on the mixing time. In our model, we assume that the transition matrix of stock volatility is typically aperiodic, reversible, and irreducible, as these properties are commonly observed in real-world financial markets.

\subsection{Portfolio Methods}

This study evaluates four distinct portfolio allocation strategies: equally-weighted investment, minimum variance, equal risk contribution, and maximum diversification. Each method is mathematically defined and applied to the ten market states identified through the K-means clustering process.

\subsubsection{Equally-Weighted Investment}

The equally-weighted investment strategy allocates an equal proportion of the total investment to each asset in the portfolio. Mathematically, if there are \( n \) assets in the portfolio, the weight \( w_i \) for each asset \( i \) is given by:

\begin{equation}
w_i = \frac{1}{n} \quad \text{for } i = 1, 2, \ldots, n.
\end{equation}

This strategy does not require any estimation of parameters and is simple to implement.

\subsubsection{Minimum Variance Portfolio}

The Minimum Variance Portfolio aims to minimize the overall variance of the portfolio, thereby reducing risk. This approach is particularly useful in creating a portfolio with the lowest possible volatility given a set of assets and their respective covariances.

Let \(\mathbf{w}\) be the vector of portfolio weights and \(\mathbf{\Sigma}\) be the covariance matrix of asset returns. The variance of the portfolio \(\sigma_p^2\) can be expressed as:

\begin{equation}
\sigma_p^2 = \mathbf{w}^T \mathbf{\Sigma} \mathbf{w}.
\end{equation}

The objective of the Minimum Variance Portfolio is to find the weight vector \(\mathbf{w}\) that minimizes \(\sigma_p^2\) subject to the constraint that the weights sum up to one. This can be formulated as the following optimization problem:

\begin{align}
\min_{\mathbf{w}} & \quad \mathbf{w}^T \mathbf{\Sigma} \mathbf{w}, \\
\text{subject to} & \quad \sum_{i=1}^{n} w_i = 1, \\
& \quad w_i \geq 0 \quad \forall i.
\end{align}

where \( n \) is the number of assets in the portfolio.

To solve this optimization problem, we can use quadratic programming techniques. The Lagrangian function for this problem is the following:

\begin{equation}
\mathcal{L}(\mathbf{w}, \lambda) = \mathbf{w}^T \mathbf{\Sigma} \mathbf{w} + \lambda \left( \sum_{i=1}^{n} w_i - 1 \right).
\end{equation}

where \(\lambda\) is the Lagrange multiplier associated with the equality constraint. 

Taking the partial derivative of the Lagrangian with respect to \(\mathbf{w}\) and setting it to zero gives:

\begin{equation}
\frac{\partial \mathcal{L}}{\partial \mathbf{w}} = 2\mathbf{\Sigma} \mathbf{w} + \lambda \mathbf{1} = 0.
\end{equation}

Solving for \(\mathbf{w}\) yields the following:

\begin{equation}
\mathbf{w} = -\frac{\lambda}{2} \mathbf{\Sigma}^{-1} \mathbf{1}.
\end{equation}

To satisfy the constraint \(\sum_{i=1}^{n} w_i = 1\), we multiply both sides by \(\mathbf{1}^T\):

\begin{equation}
\mathbf{1}^T \mathbf{w} = \mathbf{1}^T \left( -\frac{\lambda}{2} \mathbf{\Sigma}^{-1} \mathbf{1} \right) = 1.
\end{equation}

Solving for \(\lambda\):

\begin{equation}
\lambda = -\frac{2}{\mathbf{1}^T \mathbf{\Sigma}^{-1} \mathbf{1}}.
\end{equation}

Substituting \(\lambda\) back into the expression for \(\mathbf{w}\):

\begin{equation}
\mathbf{w} = \frac{\mathbf{\Sigma}^{-1} \mathbf{1}}{\mathbf{1}^T \mathbf{\Sigma}^{-1} \mathbf{1}}.
\end{equation}

This gives the optimal weights for the Minimum-Variance Portfolio. In practical applications, numerical optimization techniques such as Sequential Least Squares Programming (SLSQP) are often used to solve this problem, particularly when dealing with large numbers of assets and more complex constraints.

\subsubsection{Maximum Diversification}

The Maximum Diversification strategy aims to maximize the diversification ratio of a portfolio. The diversification ratio ($\mathrm{DR}$) is defined as the ratio of the weighted average of the volatilities of individual assets to the volatility of the portfolio. Mathematically, the diversification ratio is given by:

\begin{equation}
\mathrm{DR}(\mathbf{w}) = \frac{\sum_{i=1}^{N} w_i \sigma_i}{\sqrt{\mathbf{w}^T \mathbf{\Sigma} \mathbf{w}}},
\end{equation}

where $\mathbf{w}$ is the weight vector of the portfolio, $N$ is the number of assets, $w_i$ is the weight of the asset $i$ in the portfolio, $\sigma_i$ is the volatility of the asset $i$ and $\mathbf{\Sigma}$ is the covariance matrix of asset returns.

The objective of this strategy is to find the portfolio weights $\mathbf{w}$ that maximize $\mathrm{DR}(\mathbf{w})$:

\begin{equation}
\max_{\mathbf{w}} \left\{ \frac{\sum_{i=1}^{N} w_i \sigma_i}{\sqrt{\mathbf{w}^T \mathbf{\Sigma} \mathbf{w}}} \right\},
\end{equation}

subject to the constraints:

\begin{equation}
\sum_{i=1}^{N} w_i = 1,
\end{equation}

\begin{equation}
w_i \geq 0 \quad \forall i.
\end{equation}

This optimization problem can be solved using numerical optimization techniques such as Sequential Least Squares Programming (SLSQP).

\subsubsection{Equal Risk Contribution (ERC) Portfolio}

The Equal Risk Contribution (ERC) portfolio, often referred to as Risk Parity, aims to allocate portfolio weights so that each asset contributes equally to the overall portfolio risk. The goal is to balance the risk contributions of all assets to achieve a well-diversified portfolio.

The total risk contribution of an asset \( i \) to the portfolio is given by:

\begin{equation}
\mathrm{TRC}_i = w_i (\Sigma \mathbf{w})_i.
\end{equation}

where \(\mathrm{TRC}_i\) is the total risk contribution of asset \(i\),
     \(\mathbf{w}\) is the vector of portfolio weights,
     \(\Sigma\) is the covariance matrix of asset returns, and \((\Sigma \mathbf{w})_i\) is the \(i\)-th element of the vector obtained by multiplying the covariance matrix \(\Sigma\) by the weight vector \(\mathbf{w}\).

The objective of ERC is to equalize the total risk contributions across all assets:

\begin{equation}
\mathrm{TRC}_i = \mathrm{TRC}_j \quad \forall i, j.
\end{equation}

This can be formulated as an optimization problem where the objective is to minimize the sum of squared differences between the total risk contributions and the average risk contribution:

\begin{equation}
\min_{\mathbf{w}} \sum_{i=1}^{n} \left( \mathrm{TRC}_i - \frac{1}{n} \sum_{j=1}^{n} \mathrm{TRC}_j \right)^2,
\end{equation}

subject to the constraints:

\begin{align}
\sum_{i=1}^{n} w_i = 1, \\
w_i \geq 0 \quad \forall i.
\end{align}

The target total risk contribution for each asset in an ERC portfolio is \(\frac{\sigma_p}{n}\) of the total portfolio risk, where \(\sigma_p\) is the portfolio standard deviation.

The ERC portfolio can be implemented using numerical optimization techniques such as Sequential Least Squares Programming (SLSQP). The steps involved include:
1. Calculating the covariance matrix \(\Sigma\),
2. Defining the objective function to minimize the differences in total risk contributions,
3. Applying constraints to ensure the weights sum to one and are non-negative.

\subsection{Performance Evaluation Methods}

To determine the results of the final test, we employed several performance evaluation criteria. Firstly, we calculate the daily return and the investment value of an initial investment \$1, allowing us to graph the performance of the portfolio over time.

Next, we calculate the annual return of the portfolio over the test period, with higher values indicating better performance. This metric provided insight into the portfolio’s ability to generate returns on an annual basis.

Secondly, we assessed the volatility of the portfolio, measured as the standard deviation of portfolio returns. Lower volatility values indicated greater stability, highlighting the portfolio’s ability to maintain consistent performance without significant fluctuations.

In addition, we evaluated the Sharpe ratio, which measures the risk-adjusted return of the portfolio. A higher Sharpe ratio signifies better performance relative to the amount of risk taken, making it a crucial metric to compare different investment strategies.

Lastly, we examine the total return, volatility, and Sharpe ratio of the portfolio over the entire period. These comprehensive metrics provided an overall assessment of portfolio performance, allowing us to gauge its effectiveness in generating returns, maintaining stability, and optimizing risk-adjusted performance throughout the study.

In the following section, we will empirically implement the described methodology by first applying the K-means clustering algorithm to segment the market into ten distinct volatility-based states for two different assets. We will then construct a Bayesian Markov transition matrix to capture the transition probabilities between these states for each asset. Each state will be analyzed to determine the best-performing portfolio method: equally-weighted investment, minimum variance, equal risk contribution, or maximum diversification. The dynamic portfolio strategy will be constructed using these state-specific methods and the transition probabilities. Its performance will be evaluated in terms of annual return, annualized volatility, and Sharpe ratio, both annually and over the respective total periods for the two assets: 19 years for the first asset and 9 years for the second asset.

\section{Empirical Implementation and Results}\label{sec:empirical}

For our empirical analysis, we used daily adjusted closing prices from 11 major companies spanning from June 20, 2005, to June 20, 2024. These companies represent the top stocks of 11 sectors of the S\&P 500 as of June 20, 2024. The tickers of these companies include Apple Inc. (AAPL), Eli Lilly and Co. (LLY), JPMorgan Chase \& Co. (JPM), Amazon.com Inc. (AMZN), Alphabet Inc. (GOOGL), United Parcel Service, Inc. (UPS), Procter \& Gamble Co. (PG), Exxon Mobil Corp. (XOM), NextEra Energy Inc. (NEE), American Tower Corp. (AMT), and Linde PLC (LIN). 

For the second asset set, we used the adjusted closing prices of NASDAQ, SPY, Bitcoin, Gold, and the iShares 20+ Year Treasury Bond ETF (TLT) spanning from January 6, 2015, to June 20, 2024, as Bitcoin was listed on the market later in 2014.

We selected the first set of assets because these 11 companies collectively represent the market while maintaining simplicity. The second set of assets was chosen to represent five different types of investments, providing a diverse portfolio for our analysis.

\subsection{Implementation}

\subsubsection{Market State Classification}

We employed the K-means clustering algorithm to divide the market into 10 distinct states based on the portfolio data for each asset set. Each state was evaluated to determine the best investment method using various portfolio optimization strategies: Equal Risk Contribution (ERC), Minimum Variance (Min\_Var), Maximum Diversification (Max\_Div), and Equal Investment.

\begin{figure}[h]
    \centering
    \includegraphics[width=0.8\textwidth]{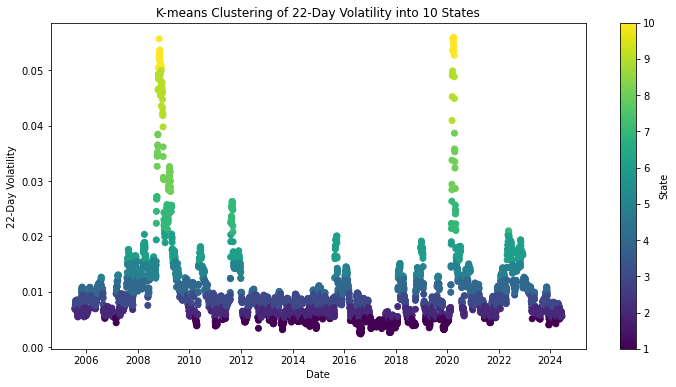}
    \caption{K-means Clustering of 22-Day Volatility into 10 States for SPY Top 11 Portfolio}
    \label{fig:kmeans_clustering1}
\end{figure}

\begin{figure}[H]
    \centering
    \includegraphics[width=0.8\textwidth]{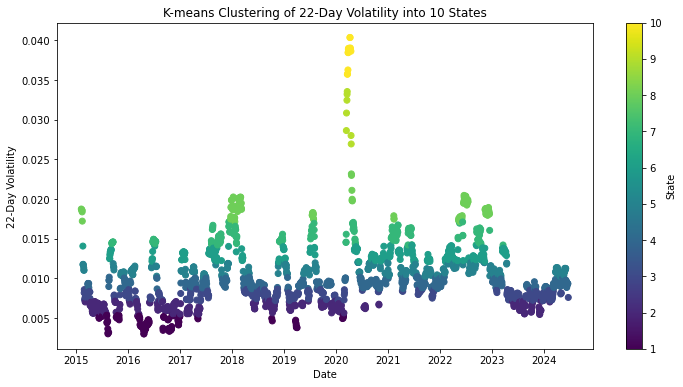}
    \caption{K-means Clustering of 22-Day Volatility into 10 States for Second Asset Portfolio}
    \label{fig:kmeans_clustering2}
\end{figure}

\subsubsection{Investment Strategies}

The investment strategies in our study were implemented using four distinct methods: equal risk contribution (ERC), minimum variance (Min Var), maximum diversification (Max Div), and equal investment. The ERC strategy allocated portfolio weights so that each asset contributed equally to the overall portfolio risk, ensuring a balanced risk distribution. The Min Var strategy focused on minimizing the overall variance of the portfolio by carefully adjusting the weights of each asset to achieve the lowest possible volatility. The Max Div strategy aimed to maximize the diversification ratio, allocating weights to enhance the diversification benefits within the portfolio. Lastly, the Equal Investment strategy allocated equal weights to all assets in the portfolio, based solely on the number of assets, regardless of their individual characteristics.

In our analysis, we calculated the total return, volatility, and Sharpe ratio for each method in different market states. Portfolio weights for the ERC, Min Var, Max Div, and Equal Investment strategies were determined using data from the entire analysis period, ensuring consistency in the application of each strategy. However, the final weights in the dynamic portfolio were adjusted based on the Bayesian Markov switching model, which provided different probabilities for each market state. Each state had a designated best-performing method, and the dynamic portfolio adjusted its weights accordingly to reflect these state-dependent probabilities. This comprehensive evaluation allowed us to understand the performance of each investment strategy under varying market conditions and identify the most effective approaches for different volatility regimes. In practice, the weights can be adjusted every 3-5 years to account for changes in market conditions and maintain optimal performance.

\subsubsection{Bayesian Markov Transition Matrix}

To model the dynamic nature of the market states, we employed a Bayesian Markov transition matrix. This transition matrix was constructed using Bayesian estimation techniques, incorporating frequency counts of state transitions to estimate the probabilities of moving from one state to another based on historical state sequences. This method enables us to predict the probability distribution of future states given the current state, ensuring that the states never have a probability of zero. 

As an example, the transition matrix for the first asset set is shown below.

\begin{figure}[H]
    \centering
    \resizebox{\textwidth}{!}{%
    \begin{tabular}{|c|c|c|c|c|c|c|c|c|c|c|}
    \hline
    State & 1 & 2 & 3 & 4 & 5 & 6 & 7 & 8 & 9 & 10 \\ \hline
    1     & 0.901227 & 0.082535 & 0.004582 & 0.002922 & 0.001539 & 0.001453 & 0.001407 & 0.001503 & 0.001421 & 0.001412 \\ \hline
    2     & 0.051824 & 0.881939 & 0.057589 & 0.003450 & 0.000896 & 0.000824 & 0.000859 & 0.000853 & 0.000861 & 0.000906 \\ \hline
    3     & 0.000995 & 0.064860 & 0.878248 & 0.049259 & 0.001931 & 0.000945 & 0.000924 & 0.000939 & 0.000965 & 0.000934 \\ \hline
    4     & 0.001344 & 0.005662 & 0.075556 & 0.851285 & 0.057849 & 0.002762 & 0.001353 & 0.001405 & 0.001404 & 0.001380 \\ \hline
    5     & 0.002061 & 0.002198 & 0.004406 & 0.086078 & 0.822895 & 0.071468 & 0.004461 & 0.002148 & 0.002155 & 0.002128 \\ \hline
    6     & 0.002859 & 0.002837 & 0.002624 & 0.005663 & 0.089236 & 0.815854 & 0.072525 & 0.002801 & 0.002848 & 0.002753 \\ \hline
    7     & 0.004482 & 0.004512 & 0.004702 & 0.008663 & 0.008616 & 0.109151 & 0.829669 & 0.021640 & 0.004222 & 0.004344 \\ \hline
    8     & 0.010154 & 0.011033 & 0.010089 & 0.010173 & 0.009721 & 0.010809 & 0.051897 & 0.825123 & 0.051001 & 0.009999 \\ \hline
    9     & 0.016267 & 0.018014 & 0.016584 & 0.018268 & 0.017042 & 0.015767 & 0.017334 & 0.086261 & 0.741515 & 0.052949 \\ \hline
    10    & 0.012331 & 0.012320 & 0.012675 & 0.012442 & 0.012099 & 0.011640 & 0.012236 & 0.012242 & 0.036882 & 0.865132 \\ \hline
    \end{tabular}%
    }
    \caption{Bayesian Markov Transition Matrix for the First Asset Set}
    \label{fig:transition_matrix_first_asset}
\end{figure}

The trend observed in this transition matrix indicates that the states tend to stick together around the diagonal. This suggests a high probability of the market remaining in the same state or transitioning to adjacent states. This behavior reflects the persistence of volatility regimes, in which the market is likely to stay in a particular volatility state or move to a state with similar characteristics rather than making abrupt transitions to vastly different states. This insight is crucial to predict future market conditions and adjust portfolio strategies accordingly.

\subsubsection{Total Return Weights Calculation}

To compute the total return weights for each investment method in each state, we employed the Bayesian Markov transition matrix in conjunction with vectors representing the optimal return methods in terms of cumulative return for each state.

For the first asset set, the optimal return methods for each state were identified as follows: Min\_Var, Min\_Var, Min\_Var, Equal, ERC, Min\_Var, Min\_Var, Equal, Max\_Div, and ERC. The binary vectors for each method were defined as:

\begin{align*}
\mathbf{p}_{\text{ERC}} &= [0, 0, 0, 0, 1, 0, 0, 0, 0, 1]^\top \\
\mathbf{p}_{\text{Min\_Var}} &= [1, 1, 1, 0, 0, 0, 1, 1, 0, 0]^\top \\
\mathbf{p}_{\text{Max\_Div}} &= [0, 0, 0, 0, 0, 0, 0, 0, 0, 1]^\top \\
\mathbf{p}_{\text{Equal}} &= [0, 0, 0, 1, 0, 0, 0, 1, 0, 0]^\top
\end{align*}

The total return weights for each method were then calculated by multiplying the transition matrix by the corresponding vector:

\begin{equation}
\text{total\_return\_weights}_{\text{method}} = \text{transition\_matrix} \times \mathbf{p}_{\text{method}}
\end{equation}

This approach ensures that the portfolio dynamically adapts to changing market conditions by leveraging the most effective investment strategy for each state. The computed total return weights reflect the probability-weighted allocation based on state transitions, allowing the portfolio to optimize returns by using the best-performing strategy in each market state.

The total return weights calculated for each method for the first asset set were as follows:

\begin{table}[H]
\centering
\caption{Total Return Weights for Each Method for the First Asset Set}
\begin{tabular}{|c|c|c|c|c|}
\hline
State & ERC     & Min\_Var & Max\_Div & Equal \\ \hline
1     & 0.002951 & 0.991204 & 0.001421 & 0.004425 \\ \hline
2     & 0.001802 & 0.993035 & 0.000861 & 0.004303 \\ \hline
3     & 0.002865 & 0.945972 & 0.000965 & 0.050198 \\ \hline
4     & 0.059229 & 0.086677 & 0.001404 & 0.852690 \\ \hline
5     & 0.825023 & 0.084594 & 0.002155 & 0.088226 \\ \hline
6     & 0.091989 & 0.896699 & 0.002848 & 0.008464 \\ \hline
7     & 0.012960 & 0.952516 & 0.004222 & 0.030303 \\ \hline
8     & 0.019720 & 0.093982 & 0.051001 & 0.835296 \\ \hline
9     & 0.069991 & 0.083966 & 0.741515 & 0.104529 \\ \hline
10    & 0.877231 & 0.061202 & 0.036882 & 0.024684 \\ \hline
\end{tabular}
\end{table}

For the second asset set:

\begin{table}[H]
\centering
\caption{Total Return Weights for Each Method for the Second Asset Set}
\begin{tabular}{|c|c|c|c|c|}
\hline
State & ERC     & Min\_Var & Max\_Div & Equal \\ \hline
1     & 0.111107 & 0.875728 & 0.013165 & 0.0 \\ \hline
2     & 0.934646 & 0.060181 & 0.005174 & 0.0 \\ \hline
3     & 0.880202 & 0.115055 & 0.004745 & 0.0 \\ \hline
4     & 0.113108 & 0.882302 & 0.004588 & 0.0 \\ \hline
5     & 0.097631 & 0.896719 & 0.005650 & 0.0 \\ \hline
6     & 0.824267 & 0.162231 & 0.013502 & 0.0 \\ \hline
7     & 0.078170 & 0.873327 & 0.048503 & 0.0 \\ \hline
8     & 0.045580 & 0.098265 & 0.856155 & 0.0 \\ \hline
9     & 0.173494 & 0.356993 & 0.469512 & 0.0 \\ \hline
10    & 0.110810 & 0.780146 & 0.109044 & 0.0 \\ \hline
\end{tabular}
\end{table}

\subsubsection{Dynamic Portfolio Construction and Performance Testing}

The dynamic portfolio strategy was constructed by dynamically allocating weights to different investment methods based on the state probabilities of the Markov transition matrix. We applied the dynamic portfolio strategy to predict performance at \(t+1\), rather than at \(t\). This approach leverages the Markov transition probabilities to better predict future states and thus optimize the portfolio's performance.

The dynamic portfolio metrics were calculated by iterating through the portfolio data to determine the state at each time point and then computing the returns for the next day based on the total return weights and the best return methods for each state. For each method (ERC, Min\_Var, Max\_Div, Equal Investment, and Dynamic), the daily returns were calculated, which were then used to compute the annualized return, volatility, and Sharpe ratio.

The cumulative return was derived using the compounded return method over the entire period: 2005 to 2024 for the first asset set and 2015 to 2024 for the second asset set. Annualized volatility was calculated from the standard deviation of daily returns, and the Sharpe ratio was computed by dividing the annualized return by the annualized volatility.

For the entire period, an initial investment of 1 dollar in each method would grow according to the cumulative return, providing a clear comparison of the growth potential and the effectiveness of risk management of each portfolio strategy. This concise approach highlights the superior performance of the dynamic portfolio in both return and risk-adjusted metrics compared to static methods.

\subsection{Results}

\subsubsection{First Asset Set: SPY Top 11 Portfolio}

The annual performance metrics for the first asset set, SPY Top 11 Portfolio, using four static methods and our dynamic portfolio strategy over the period from 2005 to 2024 are presented in Table~\ref{tab:first_asset_metrics}.

\begin{table}[H]
\centering
\caption{Annual Return for the First Asset Set}
\label{tab:first_asset_metrics}
\resizebox{\textwidth}{!}{%
\begin{tabular}{lrrrrr}
\hline
Year & ERC\_Returns & Min\_Var\_Returns & Max\_Div\_Returns & Equal\_Investment\_Returns & Dynamic\_Returns \\ 
\hline
2005 & 0.097404 & 0.139311 & 0.077847 & 0.111806 & 0.143028 \\
2006 & 0.477485 & 0.513948 & 0.453308 & 0.450023 & 0.524057 \\
2007 & 0.064272 & 0.169116 & 0.104562 & 0.125696 & 0.045541 \\
2008 & -0.178923 & -0.220533 & -0.179316 & -0.195571 & -0.209268 \\
2009 & 0.331363 & 0.531398 & 0.361124 & 0.369131 & 0.502767 \\
2010 & 0.211862 & 0.169395 & 0.204967 & 0.193348 & 0.210507 \\
2011 & 0.336085 & 0.534892 & 0.360398 & 0.352385 & 0.524274 \\
2012 & 0.071344 & -0.093848 & 0.048998 & 0.048365 & -0.017343 \\
2013 & 0.269388 & 0.378619 & 0.286913 & 0.282343 & 0.352813 \\
2014 & 0.234083 & 0.289384 & 0.242142 & 0.214865 & 0.233322 \\
2015 & 0.130046 & -0.082844 & 0.112979 & 0.080021 & 0.021170 \\
2016 & 0.355798 & 0.418460 & 0.362301 & 0.370539 & 0.413142 \\
2017 & 0.399157 & 0.280449 & 0.394685 & 0.362975 & 0.281191 \\
2018 & 0.153207 & 0.146165 & 0.138449 & 0.137082 & 0.167848 \\
2019 & 0.401882 & 0.560431 & 0.435387 & 0.423579 & 0.523706 \\
2020 & 0.341909 & 0.418695 & 0.365222 & 0.380000 & 0.390837 \\
2021 & -0.163447 & -0.031648 & -0.151327 & -0.137441 & -0.076421 \\
2022 & 0.263803 & 0.346837 & 0.283570 & 0.278146 & 0.241115 \\
2023 & 0.331852 & 0.234892 & 0.310649 & 0.274987 & 0.243776 \\
2024 & 0.188998 & 0.150970 & 0.187157 & 0.162500 & 0.172882 \\
\hline
\end{tabular}
}
\end{table}

\begin{table}[H]
\centering
\caption{Annual Volatilities for the First Asset Set}
\label{tab:first_asset_volatilities}
\resizebox{\textwidth}{!}{%
\begin{tabular}{lrrrrr}
\hline
Year & ERC\_Volatility & Min\_Var\_Volatility & Max\_Div\_Volatility & Equal\_Investment\_Volatility & Dynamic\_Volatility \\ 
\hline
2005 & 0.008678 & 0.009843 & 0.008418 & 0.008792 & 0.009049 \\
2006 & 0.008148 & 0.008800 & 0.007946 & 0.008079 & 0.008623 \\
2007 & 0.014764 & 0.015922 & 0.014096 & 0.014523 & 0.015809 \\
2008 & 0.030253 & 0.028161 & 0.027483 & 0.028709 & 0.029433 \\
2009 & 0.012370 & 0.012768 & 0.011853 & 0.011984 & 0.012506 \\
2010 & 0.010000 & 0.010068 & 0.009791 & 0.009729 & 0.009863 \\
2011 & 0.014466 & 0.014884 & 0.014024 & 0.014107 & 0.014271 \\
2012 & 0.009284 & 0.012403 & 0.009653 & 0.009484 & 0.010242 \\
2013 & 0.008576 & 0.009296 & 0.008567 & 0.008189 & 0.009261 \\
2014 & 0.009129 & 0.010147 & 0.009267 & 0.009040 & 0.009535 \\
2015 & 0.012978 & 0.013571 & 0.012957 & 0.012662 & 0.012833 \\
2016 & 0.007526 & 0.008252 & 0.007639 & 0.007466 & 0.008066 \\
2017 & 0.010241 & 0.010411 & 0.010429 & 0.010192 & 0.010120 \\
2018 & 0.014733 & 0.014825 & 0.015228 & 0.014706 & 0.014696 \\
2019 & 0.019179 & 0.021943 & 0.019238 & 0.019601 & 0.019901 \\
2020 & 0.015477 & 0.017768 & 0.015972 & 0.015724 & 0.016265 \\
2021 & 0.016397 & 0.016295 & 0.016556 & 0.016202 & 0.015927 \\
2022 & 0.016339 & 0.016668 & 0.016573 & 0.016558 & 0.016472 \\
2023 & 0.009606 & 0.010732 & 0.009817 & 0.009619 & 0.010819 \\
2024 & 0.009062 & 0.011413 & 0.009473 & 0.009475 & 0.011190 \\
\hline
\end{tabular}
}
\end{table}

\begin{table}[H]
\centering
\caption{Annual Sharpe Ratios for the First Asset Set}
\label{tab:first_asset_sharpe}
\resizebox{\textwidth}{!}{%
\begin{tabular}{lrrrrr}
\hline
Year & ERC\_Sharpe & Min\_Var\_Sharpe & Max\_Div\_Sharpe & Equal\_Investment\_Sharpe & Dynamic\_Sharpe \\ 
\hline
2005 & 10.071659 & 13.136902 & 8.059649 & 11.579434 & 14.701396 \\
2006 & 57.371268 & 57.268745 & 55.791419 & 54.463425 & 59.615340 \\
2007 & 3.675928 & 9.993527 & 6.708491 & 7.966557 & 2.248091 \\
2008 & -6.244808 & -8.186372 & -6.888429 & -7.160476 & -7.449646 \\
2009 & 25.978954 & 40.837342 & 29.623158 & 29.966843 & 39.403275 \\
2010 & 20.185768 & 15.832373 & 19.912697 & 18.846316 & 20.329096 \\
2011 & 22.541652 & 35.265132 & 24.984848 & 24.270206 & 36.035022 \\
2012 & 6.607280 & -8.372604 & 4.040191 & 4.045298 & -2.669597 \\
2013 & 30.245351 & 39.651365 & 32.324376 & 33.255147 & 37.014991 \\
2014 & 24.546323 & 27.533208 & 25.051527 & 22.660968 & 23.421162 \\
2015 & 9.249900 & -6.841128 & 7.947653 & 5.529975 & 0.870386 \\
2016 & 45.946363 & 49.500429 & 46.118205 & 48.292406 & 49.978826 \\
2017 & 38.001602 & 25.977030 & 36.885520 & 34.631776 & 26.797764 \\
2018 & 9.720275 & 9.185143 & 8.435208 & 8.641735 & 10.740829 \\
2019 & 20.432342 & 25.084362 & 22.111615 & 21.099962 & 25.813070 \\
2020 & 21.445968 & 23.001997 & 22.240529 & 23.530656 & 23.414087 \\
2021 & -10.577873 & -2.555923 & -9.744541 & -9.100121 & -5.426010 \\
2022 & 15.534028 & 20.209031 & 16.507346 & 16.194424 & 14.030823 \\
2023 & 33.504022 & 20.955654 & 30.625165 & 27.547785 & 21.607655 \\
2024 & 19.752018 & 12.351733 & 18.702068 & 16.094777 & 14.555834 \\
\hline
\end{tabular}
}
\end{table}

\begin{table}[H]
\centering
\caption{Total Performance Metrics for the First Asset Set (2005-2024)}
\label{tab:first_asset_total_metrics}
\scriptsize
\begin{tabularx}{\textwidth}{|l|X|X|X|}
\hline
Method & Total Return (\%) & Total Volatility & Total Sharpe Ratio \\
\hline
ERC\_Returns & 38.910 & 0.1715 & 226.77 \\
Min\_Var\_Returns & 53.527 & 0.2221 & 240.98 \\
Max\_Div\_Returns & 41.481 & 0.1745 & 237.65 \\
Equal\_Investment\_Returns & 37.797 & 0.1723 & 219.36 \\
Dynamic\_Returns & 49.097 & 0.2063 & 237.90 \\
\hline
\end{tabularx}
\end{table}

The dynamic portfolio strategy for the first asset set achieved a total return of 4910\%, significantly outperforming the ERC, Maximum Diversification, and Equal Investment methods. The total Sharpe ratio for the dynamic portfolio was 237.90, also outperforming the static methods except minimum variance and highlighting its effectiveness in providing a superior risk-adjusted return. Furthermore, the total volatility for the dynamic portfolio was 0.2063, which, while higher than some static methods, demonstrates the ability of the strategy to manage risk effectively through its dynamic adjustments, balancing higher returns with manageable volatility.

\begin{figure}[H]
    \centering
    \includegraphics[width=\textwidth]{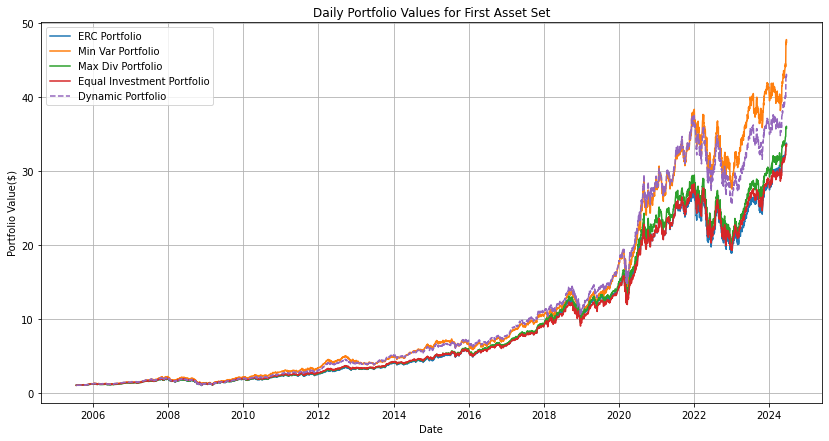}
    \caption{Investment Value Over Time for the First Asset Set with an Initial Investment of \$1}
    \label{fig:first_asset_investment_value}
\end{figure}

\begin{figure}[H]
    \centering
    \includegraphics[width=\textwidth]{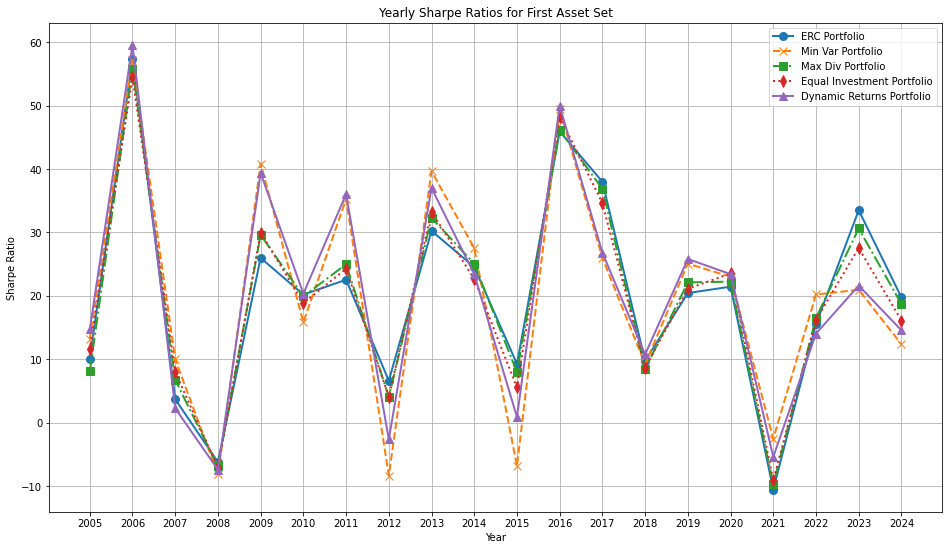}
    \caption{Yearly Sharpe Ratio for the First Asset Set}
    \label{fig:first_asset_sharpe_ratio}
\end{figure}

To illustrate the performance of the dynamic portfolio strategy, Figure~\ref{fig:first_asset_investment_value} shows the investment value over time, assuming an initial investment of \$1. Figure~\ref{fig:first_asset_sharpe_ratio} presents the yearly Sharpe ratio, highlighting the risk-adjusted returns achieved each year.

\subsubsection{Second Asset Set: NASDAQ, SPY, Bitcoin, Gold, and TLT}

The annual performance metrics for the second set of assets, including NASDAQ, SPY, Bitcoin, Gold, and TLT, over the period from 2015 to 2024 are presented in Table~\ref{tab:second_asset_metrics}. 
\begin{table}[H]
\centering
\caption{Annual Return for the Second Asset Set}
\label{tab:second_asset_metrics}
\resizebox{\textwidth}{!}{%
\begin{tabular}{lrrrrr}
\hline
Year & ERC\_Returns & Min\_Var\_Returns & Max\_Div\_Returns & Equal\_Returns & Dynamic\_Returns \\ 
\hline
2015 & 0.095424 & 0.156005 & 0.003615 & 0.096053 & 0.165819 \\
2016 & 0.715028 & 0.694940 & 0.229584 & 0.657711 & 0.740228 \\
2017 & 3.757674 & 4.218529 & 1.566528 & 3.646736 & 3.105162 \\
2018 & -0.493545 & -0.512475 & -0.370067 & -0.490234 & -0.522470 \\
2019 & 1.359878 & 1.437888 & 0.909038 & 1.347256 & 1.476886 \\
2020 & 3.321972 & 3.436156 & 2.539782 & 3.303192 & 3.168629 \\
2021 & -0.041365 & -0.044339 & -0.035539 & -0.042263 & -0.043820 \\
2022 & -0.465868 & -0.471685 & -0.437422 & -0.466479 & -0.451370 \\
2023 & 0.926852 & 0.942041 & 0.817809 & 0.926322 & 0.942852 \\
2024 & 0.454248 & 0.460317 & 0.426420 & 0.454143 & 0.455735 \\
\hline
\end{tabular}
}
\end{table}

\begin{table}[H]
\centering
\caption{Annual Volatilities for the Second Asset Set}
\label{tab:second_asset_volatilities}
\resizebox{\textwidth}{!}{%
\begin{tabular}{lrrrrr}
\hline
Year & ERC\_Volatility & Min\_Var\_Volatility & Max\_Div\_Volatility & Equal\_Volatility & Dynamic\_Volatility \\ 
\hline
2015 & 0.011259 & 0.012643 & 0.005481 & 0.010446 & 0.011539 \\
2016 & 0.015559 & 0.017942 & 0.007091 & 0.014989 & 0.015931 \\
2017 & 0.050969 & 0.053268 & 0.036969 & 0.050492 & 0.047811 \\
2018 & 0.036275 & 0.037914 & 0.026272 & 0.036004 & 0.035413 \\
2019 & 0.038888 & 0.040510 & 0.029309 & 0.038624 & 0.039072 \\
2020 & 0.046747 & 0.047361 & 0.037694 & 0.046303 & 0.045266 \\
2021 & 0.043846 & 0.044400 & 0.040480 & 0.043771 & 0.043295 \\
2022 & 0.037742 & 0.038301 & 0.033877 & 0.037723 & 0.036999 \\
2023 & 0.028875 & 0.029511 & 0.025913 & 0.028876 & 0.029304 \\
2024 & 0.036194 & 0.036666 & 0.034194 & 0.036202 & 0.036460 \\
\hline
\end{tabular}
}
\end{table}

\begin{table}[H]
\centering
\caption{Annual Sharpe Ratios for the Second Asset Set}
\label{tab:second_asset_sharpe}
\resizebox{\textwidth}{!}{%
\begin{tabular}{lrrrrr}
\hline
Year & ERC\_Sharpe & Min\_Var\_Sharpe & Max\_Div\_Sharpe & Equal\_Sharpe & Dynamic\_Sharpe \\ 
\hline
2015 & 7.587189 & 11.548392 & -1.164784 & 8.238009 & 13.503713 \\
2016 & 45.312471 & 38.175456 & 30.967458 & 43.211571 & 45.836624 \\
2017 & 73.528922 & 79.006286 & 42.103345 & 72.026459 & 64.737642 \\
2018 & -13.881292 & -13.780565 & -14.466792 & -13.893789 & -15.036143 \\
2019 & 34.711795 & 35.248085 & 30.674567 & 34.622671 & 37.542921 \\
2020 & 70.848309 & 72.340964 & 67.114508 & 71.123187 & 69.778508 \\
2021 & -1.171490 & -1.223840 & -1.124962 & -1.194022 & -1.243096 \\
2022 & -12.608595 & -12.576385 & -13.207237 & -12.630917 & -12.469772 \\
2023 & 31.752973 & 31.583360 & 31.173779 & 31.732846 & 31.833518 \\
2024 & 12.274081 & 12.281510 & 12.178239 & 12.268299 & 12.225351 \\
\hline
\end{tabular}
}
\end{table}

\begin{table}[H]
\centering
\caption{Total Performance Metrics for the Second Asset Set (2015-2024)}
\label{tab:second_asset_total_metrics}
\scriptsize
\begin{tabularx}{\textwidth}{|l|X|X|X|}
\hline
Method & Total Return (\%) & Total Volatility & Total Sharpe Ratio \\
\hline
ERC\_Returns & 65.244458 & 1.406368 & 46.385049 \\
Min\_Var\_Returns & 76.193627 & 1.520511 & 50.103948 \\
Max\_Div\_Returns & 17.967954 & 0.881578 & 20.370232 \\
Equal\_Returns & 61.222029 & 1.381796 & 44.298875 \\
Dynamic\_Returns & 59.926835 & 1.258819 & 47.597665 \\
\hline
\end{tabularx}
\end{table}

The dynamic portfolio strategy for the second set of assets achieved a notable total return of 5992.7\%, outperforming the maximum diversification method. The total Sharpe ratio for the dynamic portfolio was 47.60, indicating superior risk-adjusted performance compared to all static methods except the minimum-variance method. Furthermore, the total volatility for the dynamic portfolio was 1.2588, which is less than the volatility of the methods of equal investment, minimum variance, and equal risk contribution. This shows the effectiveness of the dynamic strategy in managing risk and adapting to market changes, providing a balanced approach to optimizing returns while maintaining lower volatility.

\begin{figure}[H]
    \centering
    \includegraphics[width=\textwidth]{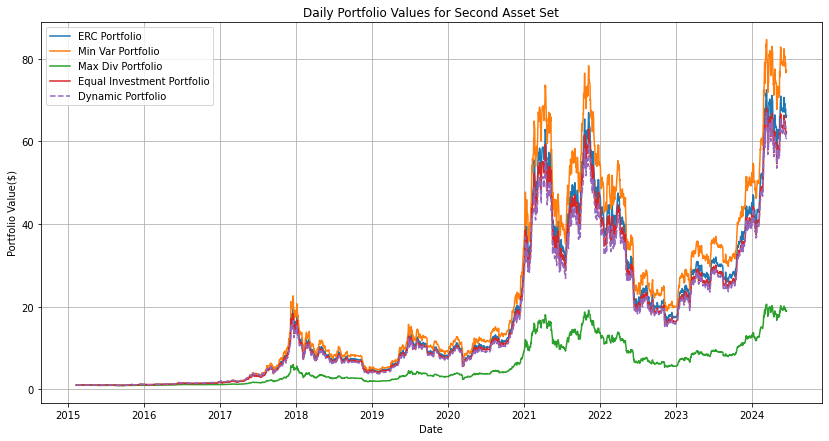}
    \caption{Investment Value Over Time for the Second Asset Set with an Initial Investment of \$1}
    \label{fig:second_asset_investment_value}
\end{figure}

\begin{figure}[H]
    \centering
    \includegraphics[width=\textwidth]{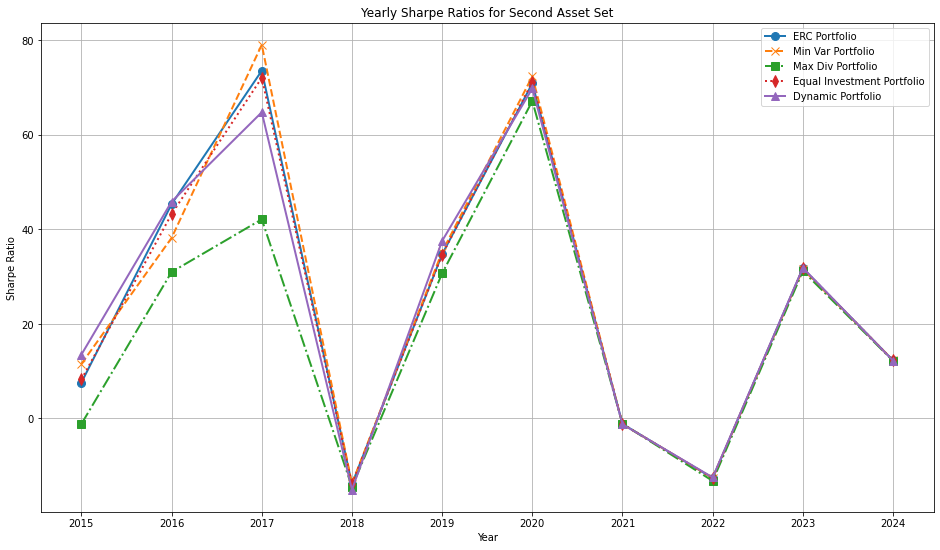}
    \caption{Yearly Sharpe Ratio for the Second Asset Set}
    \label{fig:second_asset_sharpe_ratio}
\end{figure}

To illustrate the performance of the dynamic portfolio strategy for the second set of assets, Figure~\ref{fig:second_asset_investment_value} shows the investment value over time, assuming an initial investment of \$1. Figure~\ref{fig:second_asset_sharpe_ratio} presents the yearly Sharpe ratio, highlighting the risk-adjusted returns achieved each year.

\subsection{Mixing Time of Two Assets}

In our analysis of the dynamic behavior of financial assets, we focus on computing the mixing time for two distinct assets to understand their convergence properties within a Bayesian Markov Switching framework.

\paragraph{Mixing Time Calculation}

The mixing time of a Markov chain is an important metric that indicates the number of steps required for the chain to approach its steady-state distribution. For each asset, we calculate the Second-Largest Eigenvalue Modulus (SLEM) of the transition matrix, which plays a crucial role in determining the mixing time. The key parameter used in our calculations is \(\epsilon = 0.01\), which is the threshold for proximity to the steady state.

\paragraph{First Asset Results}

For the first asset, the calculated SLEM was 0.9584, resulting in an estimated mixing time of 109 steps. This indicates that the asset's dynamics require significant iterations to stabilize, reflecting a more complex convergence process.

\paragraph{Second Asset Results}

In contrast, the second asset exhibited a SLEM of 0.9277, corresponding to a shorter mixing time of 62 steps. This suggests that the second asset's state transitions stabilize more rapidly, indicating a more straightforward convergence behavior.

\section{Discussion}\label{sec:discussion}

In this study, we explore the efficacy of various portfolio optimization methods, including Equal Risk Contribution (ERC), Minimum Variance (Min\_Var), Maximum Diversification (Max\_Div), and Equal Investment, across different market states. Utilizing a Bayesian approach to construct the Markov transition matrix, our objective was to dynamically allocate portfolio weights based on the probabilities of transitioning between these states. This approach aims to enhance future performance prediction and overall portfolio optimization by incorporating probabilistic reasoning and updating beliefs about market states as new data become available.

\subsection{Empirical Findings}

The empirical results demonstrated that the dynamic portfolio strategy, which incorporates state transitions and selects the best return methods for each state, achieved competitive performance relative to static investment strategies. Specifically, the dynamic strategy not only achieved comparable returns, but also excelled in achieving higher Sharpe ratios, particularly for more correlated asset sets. This highlights its effectiveness in managing risk and optimizing returns in a dynamic market environment.

\subsection{Portfolio 1: SPY Top 11 Portfolio}

The first asset set consisted of daily adjusted closing prices of 11 major companies from June 20, 2005, to June 20, 2024. The companies included Apple Inc. (AAPL), Eli Lilly and Co. (LLY), JPMorgan Chase \& Co. (JPM), Amazon.com Inc. (AMZN), Alphabet Inc. (GOOGL), United Parcel Service, Inc. (UPS), Procter \& Gamble Co. (PG), Exxon Mobil Corp. (XOM), NextEra Energy Inc. (NEE), American Tower Corp. (AMT), and Linde PLC (LIN).

The dynamic portfolio strategy for this asset set achieved a total return of 4910\%, significantly outperforming the methods of equal risk contribution (3891\%), maximum diversification (4148\%) and equal investment (3780\%), while being competitive with the Minimum Variance (5353\%). The annual performance comparison revealed that the dynamic portfolio outperformed the ERC in terms of returns in 10 years, Min\_Var in 10 years, Max\_Div in 11 years and Equal Investment in 13 years. This indicates the robustness and adaptability of the dynamic strategy in various market conditions.

\subsection{Portfolio 2: Second Asset Set}

The second asset set included daily adjusted closing prices of a diversified mix of assets: Nasdaq, SPY, Bitcoin, Gold, and the iShares 20+ Year Treasury Bond ETF (TLT), spanning from 2015 to 2024.

The dynamic portfolio strategy for this set of assets achieved a notable total return of 5993\%, which is higher than the maximum diversification method (1796. 79\%) but lower than the equal investment (6122\%), equal risk contribution (6524. 45\%) and minimum variation (7619. 36\%) methods. However, the dynamic portfolio's Sharpe ratio of 47.60 outperformed all static methods except for Minimum Variance, indicating superior risk-adjusted performance. The annual performance comparison highlighted that the dynamic portfolio outperformed ERC in terms of returns in 6 years, Min\_Var in 6 years, Max\_Div in 7 years, and Equal Investment in 6 years. This demonstrates the robustness and adaptability of the dynamic strategy in various market conditions.

\subsection{Analysis of Dynamic Portfolio Results}

The superior performance of the dynamic portfolio strategy can be attributed to its ability to adapt to changing market conditions by leveraging the Bayesian Markov transition matrix and dynamically allocating weights based on the best return methods for each state. This approach allows the portfolio to optimize its allocation in anticipation of future market states, rather than reacting to past performance alone.

For the first set of assets, the dynamic portfolio achieved the second-best total return of 4910\% and the second-highest Sharpe ratio of 237.90 throughout the period. This indicates that the dynamic strategy was able to deliver strong returns while maintaining superior risk-adjusted performance compared to most static methods.

For the second set of assets, the dynamic portfolio achieved a notable total return of 5993\%, outperforming the Maximum Diversification method. The total Sharpe ratio for the dynamic portfolio was 48.24, indicating superior risk-adjusted performance compared to all static methods except the minimum variance method. Furthermore, the total volatility for the dynamic portfolio was 1.2590, which is lower than the volatility of the equal investment,ERC, and equal risk contribution methods. This shows the effectiveness of the dynamic strategy in managing risk and adapting to market changes, providing a balanced approach to optimizing returns while maintaining lower volatility. Despite the dynamic portfolio's total return being the second-to-last among the methods, its Sharpe ratio was the second-best, underscoring its strong risk-adjusted performance.

The higher Sharpe ratio of the dynamic portfolio indicates that it is better prepared for state changes, thus managing risk more effectively. This strategy ensures that the best return method is maintained as much as possible while considering the state changes, providing a balanced approach to optimizing returns and managing risk.

The relatively high mixing times for both assets underscore the intricate dynamics present in their transition behaviors. A higher SLEM indicates that both assets experience slower convergence to their steady-state distributions. This necessitates a substantial volume of data to accurately capture and model the transition dynamics within a Bayesian Markov Switching framework.

These findings suggest that significant data collection and robust modeling techniques are required to effectively handle the complexity inherent in financial market transitions. The ability to accurately model these transitions is crucial for understanding the nuanced behavior of assets and making informed investment decisions in dynamic market environments.

By understanding the high data demands and convergence characteristics of these assets, we can better strategize resource allocation for data acquisition and processing, ultimately enhancing the fidelity and reliability of model outputs. This insight is crucial for optimizing the application of Bayesian Markov Switching models in financial market analysis.

Furthermore, a higher correlation among assets in the first set of assets was found to lead to better performance results, highlighting the importance of asset selection in the construction of a diversified portfolio that can achieve higher returns and better risk management.

\section{Conclusion}\label{sec:conclusion}

This study explored the efficacy of a dynamic portfolio optimization approach utilizing a Bayesian Markov transition matrix. The key findings of our analysis provide several important insights into portfolio management strategies. By dynamically allocating portfolio weights based on the probabilities of transitioning between market states, our approach aims to enhance future performance prediction and overall portfolio optimization.

The dynamic portfolio strategy demonstrated robust performance across both asset sets. For the first set of assets, it achieved a total return of 4910\%, significantly outperforming several static methods. For the second set of assets, the dynamic strategy achieved a total return of 5993\%, showing superior performance compared to the maximum diversification method, although it was slightly less than the equal investment, equal risk contribution and minimum variance methods. These results highlight the effectiveness of the dynamic strategy in adapting to market changes and optimizing returns.

In addition, the dynamic strategy consistently delivered the second highest Sharpe ratio compared to static methods for the first set of assets, indicating better risk-adjusted performance. For the second set of assets, the dynamic strategy achieved the second highest Sharpe ratio, underscoring its robustness in managing risk and adapting to state changes. Although its return was second to last.

Although promising, our study identifies several areas for future research and improvement. Incorporating additional macroeconomic and financial factors could refine state classification and enhance the predictive power of the Bayesian Markov transition matrix, improving portfolio diversification and performance.

Exploring alternative models, such as Hidden Markov Models (HMM) or Regime-Switching Models, might capture market dynamics more accurately, leading to better investment strategies. Expanding the analysis to include a broader range of assets, such as European bonds, commodities, and other cryptocurrencies, could further diversify the portfolio and improve risk management. Furthermore, incorporating more portfolio methods could help build a more sophisticated and robust dynamic portfolio, enhancing its ability to adapt to varying market conditions and optimize performance.

Developing more real-time implementation and testing frameworks is crucial to assess the practical applicability of the dynamic portfolio strategy in live trading environments. This would help evaluate the strategy's performance under actual market conditions and its responsiveness to market changes.

Several limitations should be acknowledged. The study's reliance on historical data assumes that past market behavior will repeat, which may not always hold true. The Bayesian Markov transition matrix and portfolio optimization methods are based on specific assumptions that may not capture all market dynamics and investor behavior. Additionally, the computational intensity of the dynamic strategy and the exclusion of transaction costs and other practical constraints may impact real-world performance.

In conclusion, while our study demonstrates the potential benefits of a dynamic portfolio approach, addressing its limitations and exploring the identified areas for future research could further enhance its effectiveness and applicability. The dynamic portfolio strategy shows significant promise in improving returns and sharpe ratios through better future state prediction and adaptation, but further research and refinement are needed to enhance its practical applicability across diverse market conditions.
\newpage
\bibliographystyle{plain}
\bibliography{bib}
\end{document}